\documentstyle[]{article}

\begin{document}
\setlength{\unitlength}{0.17cm}
\title{The role of phase space geometry in Heisenberg's uncertainty relation}
\author{Charis Anastopoulos \thanks{anastop@phys.uu.nl}\\
 Spinoza Instituut, Leuvenlaan 4, \\
3584HE Utrecht, The Netherlands \\ \\
and \\ \\
Ntina Savvidou \thanks{ntina@ic.ac.uk}\\
 Theoretical Physics Group, \\
Imperial College, SW7 2BZ London, \\UK } \maketitle

\renewcommand {\thesection}{\arabic{section}}
 \renewcommand {\theequation}{\thesection. \arabic{equation}}
\let \ssection = \section
\renewcommand{\section}{\setcounter{equation}{0} \ssection}

\begin{abstract}
Aiming towards a geometric description of quantum theory, we study
the coherent states-induced metric on the phase space, which
provides a geometric formulation of the Heisenberg uncertainty
relations (both the position-momentum and the time-energy ones).
The metric also distinguishes the original uncertainty relations
of Heisenberg from the  ones that are obtained from
non-commutativity of operators. Conversely, the uncertainty
relations can be written in terms of this metric only, hence they
can be formulated for any physical system, including ones with
non-trivial phase space. Moreover, the metric is a key ingredient
of the probability structure of continuous-time histories on phase
space. This fact allows a simple new proof the impossibility of
the physical manifestation of the quantum Zeno and anti-Zeno
paradoxes. Finally, we construct the coherent states for a
spinless relativistic particle, as a non-trivial example by which
we demonstrate our results.
\end{abstract}

\section{Introduction}
 In the present paper we focus on the
geometric description of quantum theory, through the study of
coherent states. Of key importance is a metric $ds^2$ on phase
space, which is induced by the coherent states construction.
Remarkably, this metric includes both types of uncertainty in its
structure, but in a distinct mathematical fashion. Moreover, the
Heisenberg uncertainty relations can be written straightforwardly
in terms of the metric as $\delta s^2 \sim 1$. This is an
important result, because it allows the formulation of the
Heisenberg uncertainty relations for any physical system, in
particular the ones with non-trivial phase space.

The uncertainty relations are a key component of quantum theory
that signifies a fundamental inability to simultaneously determine
the values of the fundamental physical quantities with arbitrary
accuracy. There exist two different versions of the uncertainty
relations. The original version is due to Heisenberg \cite{He27};
it refers to measurements of individual quantum systems and does
not employ the mathematical formalism of quantum theory, but
derives the uncertainty relations as a consequence of
wave-particle duality. The second version employs the formalism of
quantum theory and derives the uncertainty relation as a
consequence of the non-commutativity of operators. However, this
derivation does not refer to individual quantum systems, as the
uncertainties are, essentially, the mean deviations of
measurements in a statistical ensemble.

This distinction is not usually emphasised, mainly because the
statistical uncertainty relation, which is mathematically
concrete, can be viewed as a consequence of the Heisenberg's
uncertainty relations. However, the distinction persists not only
because of the different context of the two types of uncertainty,
but also because the time-energy uncertainty relation does not
have an operator analogue \footnote{The reason for this is the
inability to define a time operator. However, one can derive some
operator versions of the time-energy uncertainty relation by
considering some quantum mechanical variable as a clock that
measures time (see \cite{Bus01} for a recent review). However, no
quantum mechanical clock is guaranteed to always move forward in
time \cite{UnWa89}.}.

Our discussion is closely related to the consistent histories
description of quantum theory \cite{Gri84, Omn8894, GeHa9093,
Har93a, I94}. We show that the probability assignment for
continuous-time histories on phase space depends explicitly  on
the  metric $ds^2$, a fact that highlights the geometric character
of quantum probability. This is intended to provide a starting
point for a procedure of quantisation in terms of histories that
will be fully geometric. Moreover, the relation of the metric to
the uncertainty relation allows us to demonstrate in a simple
manner that the quantum Zeno effect \cite{MiSu77} cannot be
physically manifested.

The structure of this paper is the following. In section 2 we give
the relevant background, namely the formalism of quantum
mechanical histories and the geometry of coherent states. The main
results of the paper are in section 3. We show how the
probabilities for continuous-time histories have geometrical
origin, the relation of the metric to the uncertainty relations
and their consequences. In section 4 we give an explicit example
for our results, namely we construct the coherent states for a
spinless relativistic particle, identify the corresponding
geometry and consequently write the relativistic version of the
Heisenberg  uncertainty relations.

\section{Background}
\subsection{Quantum mechanical histories}

 The histories formalism describes quantum mechanical
systems in terms of the properties that refer to more than one
moment of time. These are the {\em histories}. In this sense the
relation of the standard quantum mechanical formalism to the
histories one is analogous to the relation of Hamilton's
formulation vs the Lagrangian action one in classical mechanics
\cite{Sav99a, Sav01b}: Hamilton's formulation is based on the
properties of a system at a single moment of time and studies
their evolution, while the action principle starts with paths
(histories) and seeks which one of them are realisable. A similar
distinction can be made in the    context of classical probability
theory: the evolution of single-time properties is effected by
 the Fokker-Planck equation for a probability distributions,
 while the theory of stochastic processes
deals with the probabilities of paths. The formalism of quantum
mechanical histories can be developed in such a way as to
explicitly denote the analogy with both classical mechanics
\cite{Sav99a} and stochastic processes \cite{Ana03, Ana02b}.

The histories description arose as part of a particular
interpretation of quantum theory, namely consistent or decoherent
histories  \cite{Gri84, Omn8894, GeHa9093, Har93a, I94}. This
interpretation considers quantum theory as a theory that describes
{\em individual closed} systems, in contrast with the Copenhagen
interpretation that refers to systems in the context of a
measurement procedure and interprets the probabilities in terms of
ensembles of quantum systems.

A key feature of the histories description is that the
probabilities  for histories do not satisfy the additivity
condition of Kolmogorov. If $\alpha$ is a proposition about a
history and $\bar{\alpha}$ its negation, it is not necessary that
$p(\alpha) + p(\bar{\alpha}) = 1$. Hence even if we can ascertain
that $p(\alpha) = 1$, we cannot preclude the possibility that
$\bar{\alpha}$ can also take place. This would not be a problem
for a theory that cares to describe only measurement outcomes as
the measurement of history $\alpha$ and the measurement of history
$\bar{\alpha}$  refer to different experimental setups. However,
it is quite problematic for the predictability of any theory that
describes individual systems. The consistent or decoherent
histories interpretation avoids this problem by postulating that
one can use probabilities to make inference only when one works
within particular sets of histories within which the probabilities
are additive. Such sets are known as consistent sets and they are
obtained by a decoherence condition.

The consistent histories interpretation has its points of
contention, namely that one can make inferences within different
consistent sets and conclude contrary properties \cite{Kent97}---
without, however, arriving at a logical contradiction
\cite{GriHa98}. Like all interpretations of quantum theory that
aim to describe individual systems, the consistent histories has
to accept that properties assigned to systems are contextual
\cite{Ish97}.

However, the history formalism exhibits a pluralism and can be
employed without a commitment to the consistent histories
interpretation.  In the present paper we will assume a convenient
Copenhagen stance, which states that {\em probabilities refer to
experiments} and are determined by measurements of ensembles of
quantum systems.

A history is a time-ordered sequence of properties of a quantum system. A single time
property is determined by a projection operator, so a history is an ordered set of
projectors labelled by time
\begin{equation}
\alpha \rightarrow (\hat{\alpha}_{t_1}, \ldots, \hat{\alpha}_{t_n})
\end{equation}
To each history one can assign the operator $\hat{C}_{\alpha}$ defined by
\begin{equation}
\alpha \rightarrow \hat{C}_{\alpha} = \hat{\alpha}_{t_1}(t_1) \ldots
\hat{\alpha}_{t_n}(t_n),
\end{equation}
where $\hat{\alpha}(t)$ refers to the Heisenberg picture projector
\begin{equation}
\hat{\alpha}_t(t) = e^{i \hat{H}t} \hat{\alpha}_t e^{-i\hat{H}t}.
\end{equation}
 Note the distinct appearance of time $t$ as an argument of the temporal
 ordering of projectors and
as the parameter of Heisenberg time evolution \cite{Sav99a,
Sav99b}. Here $\alpha_t$ is a Schr\"odinger picture operator at a
fixed time $t$; in general it is different from the time-parameter
$t$ that appears in $e^{-i \hat{H}t}$. This is a basic feature of
the temporal structure of histories quantum theory as has been
identified in previous work of one of us \cite{Sav99a, Sav01b}. In
particular,  we employ the word time to denote two different
physical concepts in physical theories. The first is the notion of
causality in spacetime, i.e. the designation of whether one event
is prior or later to another. The other is the notion of
evolution, i.e. time as a parameter in the equations of motion
that measures how much a physical system has changed from some
initial instant. In classical mechanics and ---to a large
extent--- in quantum theory this distinction is not essential: the
time-parameter of Hamilton's or Heisenberg's time evolution is
usually thought of as incorporating both notions of time.

However, the distinction is natural in the histories framework and
may be essential in general relativity \cite{Sav01b, Sav03} as
 any attempt to quantise this theory needs to deal with the ``problem of
time'' \footnote{The problem of time amounts to the following. In
General Relativity the Hamiltonian vanishes due to constraints.
How can we reconcile this with the causal description necessary
for any physical theory? This  is a problem already at the
classical level, but at the quantum level it is even more acute.
Because standard quantum theory necessitates a background causal
structure in order to make sense, while in General Relativity the
causal structure is itself a dynamical quantity.}(see \cite{Ish92,
Kuc91} for a review).

From the operators (2.2) one can construct the {\em decoherence
functional}. This is a complex valued function of pairs of
histories
\begin{equation}
(\alpha, \beta) \rightarrow d(\alpha, \beta) = Tr \left( \hat{C}^{\dagger}_{\alpha}
\hat{\rho}_0 \hat{C}_{\beta} \right).
\end{equation}
Here $\hat{\rho}_0$ is the density matrix at time $t = 0$. This
object contains all information related to the state of the system
and the dynamics. Its diagonal elements are  the probabilities
$p(\alpha)$ for the histories,
\begin{equation}
d(\alpha, \alpha) = p(\alpha).
\end{equation}
Its off-diagonal elements have an interpretation in terms of {\em
geometric phases} \cite{Ber84, WilS89, Sim83}. In the generic case
$d(\alpha, \beta)$ is a complex number $ r e^{i \theta}$. The
phase $e^{i \theta}$ is a special case of the Pancharatnam  phase
\cite{Pan56, AnaAh87} and is, in principle, measurable
 \cite{Ana03}. The modulus $r$ can also be operationally determined, but
is less interesting \footnote{ The decoherence functional plays an
additional role . If  in place of the projectors in equations
(2.3) and (2.4) we insert a self-adjoint operator $\hat{A}$ then
the values of the decoherence functional are the mixed
time-ordered and anti-time-ordered correlation functions of
$\hat{A}$. Thus the decoherence functional can be identified
\cite{An01b, Ana03} as the object that contains all temporal
correlation functions of the theory: it is known as the
closed-time-path generating functional and was introduced by
Schwinger \cite{Schw61}.}.

It is easier to see the appearance of a geometric phase in the
off-diagonal elements of the decoherence functional in the
particular case of very fine-grained histories, i.e. histories
consisting of one-dimensional projectors \cite{AnSav02}. To this
end we consider two histories \\
$\alpha = (|\psi_{t_1} \rangle
\langle \psi_{t_1} |, \ldots |\psi_{t_n} \rangle \langle
\psi_{t_n} |)$ and $\beta = (|\psi'_{t'_1} \rangle \langle
\psi'_{t'_1} |, \ldots |\psi_{t'_m} \rangle \langle \psi_{t'_m}
|)$ and we construct the decoherence functional.

For simplicity, we take $\hat{H} = 0$ and $\hat{\rho}_0$
corresponding to a pure state $\psi_{0}$. From (2.4) we get
\begin{equation}
 d(\alpha, \beta) = \langle \psi_{t_n}|\psi_{t_{n-1}} \rangle \ldots
\langle \psi_{t_1}|\psi_0 \rangle \langle \psi_0| \psi'_{t'_1} \rangle \ldots \langle
\psi'_{t'_{m-1}}| \psi'_{t'_m} \rangle
\end{equation}
The right-hand side of this equation is known as the ${\em n+m+1}$
{\em Bargmann invariant}\footnote{ The invariance it refers to is
with respect to the group $U(1)^{n+m+1}$ in its action $ |\psi_t
\rangle \rightarrow e^{i \alpha_t} |\psi_t \rangle$.}. Moreover,
if we take the continuum limit defined by  $\delta t = \sup \{|t_i
- t_{i-1}|, |t'_j - t'_{j-1}| \}$ going to zero, the decoherence
functional reads
\begin{equation}
d(\alpha, \beta) = e^{\left( \sum_{i=1}^n \langle \psi_{t_i }|
\psi_{t_i} - \psi_{t_{i-1}} \rangle- \sum_{i=0}^m \langle
\psi_{t'_i}| \psi_{t'_i} - \psi_{t'_{i-1}} \rangle \right)} +
O(\delta t^2).
\end{equation}
This converges to
\begin{equation}
d(\alpha, \beta) = e^{i \int_C A},
\end{equation}
where $A$ is an one-form defined over the projective Hilbert space
$P{\cal H}$
\begin{equation}
i A = \langle \psi | d \psi \rangle,
\end{equation}
with $d$ the exterior derivative on $P{\cal H}$ and $C$ is the
closed loop formed by the combination of the path $ \psi(\cdot)$
and $\psi'(\cdot)$. In fact, $A$ is a connection one-form, known
as the Berry connection.

From equation (2.7) it is clear that  $d(\alpha, \beta)$ equals
the holonomy of the Berry connection over $C$: this is
 the well-known Berry phase associated to the path \cite{Sim83}.
 Its value is invariant under the gauge $U(1)$
transformation (we suppress the primes for simplicity)
\begin{equation}
|\psi_t \rangle \rightarrow e^{i \theta(t)} | \psi_t \rangle.
\end{equation}

\subsection{Phase space in quantum theory}

The  expressions (2.6) above refer, in general, to histories on a
Hilbert space ${\cal H}$. In what follows, we will focus on
histories that correspond to points of the {\em classical} phase
space  and we shall translate all structures we encountered so far
in that context.

The reason for this is primarily interpretational. In particular
\cite{An01a, Ana03} it has been proposed in \cite{An01a,
Ana03}that quantum theory can be formulated solely in the
classical phase space and without making any reference to the
quantum mechanical Hilbert space. In this construction  the
fundamental observables are {\em commutative} and the quantum
behavior is contained in the relative phases between histories.
This information is fully encoded in the decoherence functional.
It was shown that all predictions of quantum theory can thus be
derived, including the predictions associated to Bell's theorem.

The key result from this is the important statement that
non-commutativity is not an essential feature of quantum theory,
in the sense that predictions identical to the ones of tandard
quantum theory can be obtained from an axiomatic scheme, which is
fundamentally based on commutative variables. This fact is rather
important as it provides a new perspective towards possible
interpretations of quantum theory (see \cite{Ana03} for a
summary).

In this approach  a reconstruction theorem was proven
\cite{Ana03}. If, in particular, we start from a theory that
satisfies the history axioms on phase space, we can uniquely write
a standard quantum theory on a Hilbert space, in such a way that
the statistical predictions are identical. The translation between
the phase space and the Hilbert space concepts is effected by
means of {\em coherent states}. To their description we shall now
turn.

\subsection{Coherent states and their geometry}
\paragraph{Hilbert space geometry.} In most approaches to quantum
theory the rich geometry of the Hilbert space is not particularly
emphasised, even if this geometry is a fundamental part of quantum
theory. The reason for this is that the formalism of quantum
theory, as it was originally developed, does not highlight the
geometric structure of the theory. However, this structure is
present and is manifested in the appearance of the geometric
phases.

It is well known that  the inner product of  a complex Hilbert
space ${\cal H}$ defines a metric and a symplectic form, from its
real and imaginary part respectively. If we remove the null vector
from ${\cal H}$ the resulting space is a total space of a bundle,
with base space the projective Hilbert space, fiber the complex
numbers ${\bf C}$ and projection map $\pi: | \psi \rangle
\rightarrow (\langle \psi| \psi \rangle)^{-1} | \psi \rangle
\langle \psi|$. This is clearly a line bundle, which is often
called the {\em Hopf bundle}, and denoted as {\bf H}.

Due to the bundle structure the inner product on $H$ induces  a
metric, a connection and a symplectic form \cite{Pag87} on $P
{\cal H}$. In terms of a representative normalised vector
$|\psi\rangle$,
\begin{eqnarray}
ds^2 = \langle d\psi|d \psi \rangle - |\langle \psi|d \psi \rangle|^2,\\
iA = \langle \psi| d \psi \rangle, \\
\Omega = d A.
\end{eqnarray}
The connection one-form $A$ is the  Berry connection we
encountered earlier, the symplectic form is the curvature of $A$
and the metric is known as the Fubini-Study metric. Note that a
transformation $|\psi \rangle \rightarrow e^{i \theta}
|\psi\rangle$ affects the connection one-form
\begin{equation}
A \rightarrow A + d \theta.
\end{equation}

\paragraph{General coherent states.}
The coherent states form a bridge between quantum theory and
classical symplectic mechanics. A set of coherent states is
defined as a map from a manifold $\Gamma$ to the projective
Hilbert space $P{\cal H}$,
\begin{equation}
i: z \in \Gamma \rightarrow |z \rangle \langle z| \in PH.
\end{equation}
Note that $|z \rangle$ refers to a representative (up to a phase)
normalised vector on the complex Hilbert space. Part of the
definition of coherent states is often the requirement that the
set of vectors $|z \rangle$ forms a resolution of the unity, i.e.
given a measure $\mu$ on $\Gamma$, we have
\begin{equation}
1 = \int d \mu(z) |z \rangle \langle z|
\end{equation}

Coherent states are often constructed by means of group
representations. If a group has a unitary representation
$\hat{U}(g), g \in G$ on a Hilbert space $H$, then we can
construct the vectors $\hat{U}(g) |0 \rangle$, where  $| 0
\rangle$ is a reference vector. The usual choice for $| 0 \rangle$
is either the minimum energy state or a vector that is invariant
under the maximal compact subgroup of $G$. Then we define the
equivalence relation $~$ on $G$ as follows: \\ \\
$g \sim g'$ if
there exists $e^{i \theta} \in
U(1)$ such that $\hat{U}(g) |0 \rangle = e^{i \theta} \hat{U}(g') |0 \rangle$.\\ \\
Defining the manifold $\Gamma = G/ \sim$, the map
\begin{equation}
[g] = z \in \Gamma \rightarrow \hat{U}(g) |0 \rangle \langle 0|
\hat{U}^{\dagger}(g)
\end{equation}
defines a set of coherent states $|z \rangle$. Furthermore, this set does possess a
resolution of the unity.

The  map $i:\Gamma \rightarrow P{\cal H}$  can be employed to
pull-back the geometrical objects defined on $P{\cal H}$ s to
$\Gamma$. In particular we can  define on $\Gamma$, the pull-back
bundle $i^*{\bf H}$, the metric, the connection and its curvature
form
\begin{eqnarray}
ds^2_{\Gamma} =  \langle d z|d z \rangle -  |\langle z|d z \rangle|^2,
\\
iA_{\Gamma} = \langle z| dz \rangle,
\\
\Omega_{\Gamma} = dA_{\Gamma}.
\end{eqnarray}

The two-form $\Omega_{\Gamma}$ can, in general, be degenerate: if
it is not,  $\Gamma$ has the structure of a symplectic manifold.
Then the Liouville form $\Omega \wedge \ldots \wedge \Omega$
defines a measure on $\Gamma$ and makes possible the existence of
a resolution of a unity. We want to emphasise that  the bundle
structure, the U(1) connection and the metric on $\Gamma$ are
 defined, {\em irrespective} of whether there exists a
resolution of the unity or not.

Our main motivation is to study the quantum structure of histories
on phase space and  the coherent states  are the key feature,
which translates the structures of standard quantum theory into
geometric objects on the classical phase space.

The idea that the classical phase space is a fundamental
ingredient of quantum theory is very intriguing. However the study
of coherent state histories is interesting by itself, because it
allows one to reproduce all quantum mechanical predictions. In
order to demonstrate this
  feature of coherent states we need to recall two basic results.

 Firstly, the function that gives the inner product between two coherent states $\langle
z|z' \rangle$ contains sufficient information to construct the
corresponding Hilbert space. Quite simply the inner product
$\langle z| z' \rangle$ defines the matrix elements of a projector
$E$ on the Hilbert space $L^2(\Gamma)$: the range of this
projector is the Hilbert space of the theory \cite{KlSk85}.

Secondly, most operators on ${\cal H}$ can be written in terms of
functions on $\Gamma$ as
\begin{equation}
\hat{A} = \int d \mu(z) f_A(z) |z \rangle \langle z|,
\end{equation}
provided there exists a resolution of the unity. Hence if we
construct  the decoherence functional on coherent states, we  can
reproduce its values for  any history.

\paragraph{The standard coherent states.} The most usual case
of coherent states are the ones associated with the
 Weyl group. Let us consider for simplicity the one-dimensional case,
  in which the generators are
 $\hat{x}, \hat{p}, \hat{1}$. We shall use the Schr\"odinger representation,
  in which the vectors are square integrable functions of $x$ and the elements of the
   Weyl group act as
\begin{eqnarray}
e^{-i \hat{p}q} \psi(x) = \psi(x-q) \\
e^{- i \hat{x} p} \psi(x) = e^{-ipx} \psi(x).
\end{eqnarray}

We choose a reference state $\psi_0(x)$ with vanishing expectation
value of $\hat{x}$ and $\hat{p}$. Then for $z = (q,p) $ the
coherent states read
\begin{equation}
\psi_z(x) = e^{ipx} \psi_0(x-q).
\end{equation}

We can then  show that
\begin{eqnarray}
A &=& p dq \\
 ds^2 &=& (\Delta p)^2 dq^2 + (\Delta q)^2 dp^2 + 2 C_{pq} dpdq
\end{eqnarray}
where $\Delta q$ and $\Delta p$ are the standard quantum
mechanical  uncertainties associated to the coherent state
 $|z \rangle$ (now written in the Dirac notation) and
   $C_{pq} = \frac{1}{2}\langle z|\hat{x} \hat{p} +\hat{p} \hat{x}| z \rangle -
    2 \langle z|\hat{x} |z \rangle \langle z|
\hat{p}|z \rangle$ is the correlation between position and
momentum. Note that neither the uncertainties nor the correlations
depend on $z$ and can  therefore be defined with respect to the
reference vector $\psi_0$.

We usually choose the reference vector $\psi_0$, so that $C_{pq} =
0$. The  standard choice is
\begin{equation}
\psi_0(x) = \frac{1}{( \pi \sigma^2)^{1/4}} e^{-\frac{x^2}{2 \sigma^2}}.
\end{equation}

In what follows, we shall show the important role of the geometry
of coherent states in the probability structure of continuous-time
histories.

\section{Phase space histories}

We will now study the decoherence functional for phase space
histories. For this purpose it is sufficient to examine its values
for the  finest-grained histories, as any history can be obtained
by coarse-graining on phase space \cite{Omn4, Ana03}. The
classical phase space is  a symplectic manifold $\Gamma$ of
dimension $n$, whose points will be denoted by $z$.

\subsection{Phase space histories with zero Hamiltonian}

\subsubsection{The expansion of the propagator}

We shall first consider the case of trivial dynamics, i.e.
$\hat{H} = 0 $. From equation (2.6) we see that in order to
construct the decoherence functional we need to compute
 the inner product $\langle z|z ' \rangle$.

In order to study continuous-time histories we need to consider
the case that $z'$ is infinitesimally close to $z$: $z' = z +
\delta z$. We then expand the inner product in powers of $\delta
z$ and keep terms of order $(\delta z)^2$
\begin{equation}
\langle z | z + \delta z \rangle = 1 + \langle z| \partial_i z
\rangle \delta z^i + \frac{1}{2} \langle z| \partial_i \partial_j
z \rangle \delta z^i \delta z^j + O(\delta z^3).
\end{equation}
>From equations (2.18-2.19) we see that the components of the
connection form and the metric on $\Gamma$ read

\begin{eqnarray}
i A_i(z) = \langle z|\partial_i z \rangle, \\
g_{ij}(z) = \langle \partial_i z| \partial_j z\rangle + A_i(z)
A_j(z),
\end{eqnarray}

where the indices $i, j$ run from 1 to $n$.

Hence,
\begin{equation}
\langle z|\partial_i \partial_j z \rangle = - g_{ij}(z) - A_i(z)
A_j(z) + i \partial_i A_j(z).
\end{equation}

This implies that
\begin{equation}
\langle z| z + \delta z \rangle = \exp \left( i A_i(z +
\frac{\delta z}{2}) \delta z^i - \frac{1}{2} g_{ij} \delta z^i
\delta z^i \right) + O(\delta z^3).
\end{equation}

We now consider  a pair of histories $\alpha$ (corresponding to $
r \rightarrow |z_r \rangle$) and $\beta$ (corresponding to $r'
\rightarrow |z'_{r'} \rangle$). The integers $r$ and $r'$ denote
the time instants and  take values $r = 1, \ldots n$, $r' = 1
\ldots m$. Furthermore, we  assume that the initial point on phase
space is given by
 $|z_0 \rangle$ and that $z_n = z'_m$.

We construct the loop $a \rightarrow |z_a \rangle$, with $a = 0,
n+m$ from the combination of the two previous paths, namely
\begin{eqnarray}
z_a =      \left\{ \begin{array}{cc} z_r,& \hspace{1cm} a = 1,
\ldots n
\\ z'_{a-n},& \hspace{0.5cm} a= n+1, n+m \\
  z_0,& \hspace{1cm} a = 0 \end{array} \right.
\end{eqnarray}

From equation (2.6) we  write the decoherence functional
\begin{equation}
d(\alpha, \beta) = \prod_a \langle z_a| z_a + \delta z_a \rangle,
\end{equation}
where $\delta z_a = z_{a+1} - z_a$ and where we have refrained
from writing explicitly a time parameter. Finally,
\begin{equation}
d(\alpha, \beta) = \exp \left( i \sum_a A(z_a + \frac{\delta
z_a}{2}) \delta z_a - \frac{1}{2} \sum_a \delta s_a^2 \right)  +
O(\delta z^3).
\end{equation}

If  $|\delta z_a^i| < \epsilon$ and we subsequently let $\epsilon
\rightarrow 0$ \footnote{This means that the paths are continuous
and their variation is bounded.}, the expression (3.8) converges
to $\exp (i \int_CA)$, as we showed in section 1.

We should note here that for any fine-grained history $\alpha$,
the probability $d(\alpha, \alpha) \rightarrow 1$ in the continuum
limit. This seems to imply that
 a continuous measurement of whether the quantum system follows a path forces
 the system to actually follow the path.

This behaviour is reminiscent of the  quantum Zeno effect
\cite{MiSu77}, which refers to the following behaviour.
 When we monitor {\em continuously} a quantum system
in order to see if a transition has occurred, we inhibit the
transition so that the probability that the system remains in the
initial state is equal to one. In the present case an anti-Zeno
effect \cite{BalRo00} is manifested, a probability equal to one is
assigned to a general path rather than to the preservation of a
state in time.

\subsubsection{Uncertainty relation}
As we showed in section 1.4, the phase space metric for the case
of the Weyl group is written
\begin{equation}
 \delta s^2 = (\Delta p)^2
 \delta q^2 + (\Delta q)^2 \delta p^2 + 2 C_{pq} \delta p \delta
 q.
\end{equation}

In this equation, there exist two types of ``uncertainties'' for
physical quantities, namely $\delta $ and $\Delta$. Their physical
interpretations are distinct and for this reason we will elaborate
on the meaning of the Heisenberg uncertainty relations.

When Heisenberg first wrote the uncertainty relations, he was
referring to the uncertainty in the measurement of position and
momentum of an {\em individual} quantum system. He did not make
use of the formalism of quantum mechanics; the uncertainty
relation was a consequence of solely the wave-particle duality. It
is interesting to recall his argument: In order  to measure the
position of a particle, we need to employ photons (or electrons)
of some wavelength $\lambda$. There will then be an inaccuracy in
the measurement of the order of $\delta x \sim \lambda$. In order
to measure momentum with the same source of photons, we cannot
have higher accuracy than the momentum of a single photon, so
$\delta p \sim \frac{\hbar}{\lambda}$. We altogether  have an
uncertainty $\delta x \delta p \sim \hbar$.

On the other hand, starting from the formalism of quantum
mechanics one can derive another uncertainty relation
\begin{eqnarray}
\Delta q \Delta  p \geq \frac{\hbar}{2},
\end{eqnarray}
 as a consequence of the non-commutativity of the operators. Here $\Delta q$ is defined as
\begin{eqnarray}
(\Delta q)^2 = \langle \psi| \hat{q}^2| \psi \rangle -  (\langle
\psi| \hat{q}| \psi \rangle)^2,
\end{eqnarray}
and similarly is $\Delta p$ defined. These quantities are
interpreted as mean deviations of the distributions of  position
and momenta respectively, hence they are {\em statistical
objects}. Each of them refers to a different type of measurement,
hence a different experimental set-up; {\em they do not refer to
individual systems}. In fact, they can be viewed as features of
the quantum state $|\psi \rangle$, through which they are defined.
In what follows we shall refer to the first uncertainty principle
as the Heisenberg uncertainty principle and to the second as the
statistical uncertainty principle.

A widespread belief is that the uncertainty relations  proves that
the phase space properties of physical systems are not
simultaneously definable. This is not true for either of them.
Both uncertainty relations make explicitly reference to
measurements and hence refer to open systems. The only
demonstration of such non-definability comes in the wake of the
Kochen-Specker theorem  \cite{KoSp67}, but this necessitates the
dubious assumption that the quantum mechanical formalism
(operators and states) refers to {\em individual systems} rather
than ensembles. Clearly in the Copenhagen interpretation the
question of simultaneous definability of observables corresponding
to non-commuting operators is irrelevant, since the Copenhagen
interpretation deals exclusively with measurement outcomes.

From the above analysis, it is clear that the quantities with the
$\Delta$ in equation (3.9) refer to the statistical properties of
the coherent states, while $\delta p$ and $\delta q$ refer to the
difference between  $|z \rangle$ and $|z + \delta z \rangle$. If
we interpret the decoherence functional in terms of measurements
this is  the difference between  two filters measuring the phase
space properties of the physical system.

No physical measurement
 device has infinite accuracy on phase space, because the Heisenberg
 uncertainty relation sets a bound on the sharpness we can
achieve by means of phase space measurements. Hence,
\begin{equation}
 \delta q \delta p  \sim 1.
\end{equation}

Taking this into account, the infinitesimal distance $\delta s^2$
satisfies
\begin{equation}
\delta s^2 \geq (\Delta p)^2 \delta q^2 + \frac{(\Delta q)^2}{
\delta q^2} + 2 C_{pq}.
\end{equation}

Minimising the right-hand-side with respect to $\delta q$ we
obtain
\begin{equation}
\delta s^2 \geq 2 (\Delta p \Delta q + C_{pq}).
\end{equation}

We can further simplify this expression by  using a generalised
statistical uncertainty relation \cite{DKM80}
\begin{equation}
(\Delta q)^2 (\Delta p)^2 - C_{pq}^2 \geq \frac{1}{4},
\end{equation}
which implies that
\begin{equation}
|C_{pq}| \leq \sqrt{ (\Delta q)^2 (\Delta p)^2 - \frac{1}{4}}.
\end{equation}
So we have
\begin{eqnarray}
\Delta p \Delta q + C_{pq} \geq \Delta p \Delta q - |C_{pq}| \nonumber \\
\geq \Delta p \Delta q -  \sqrt{ (\Delta q)^2 (\Delta p)^2 - \frac{1}{4}} \geq \frac{1}{2}.
\end{eqnarray}
Substituting into (3.12) we get
\begin{equation}
\delta s^2 \geq 1.
\end{equation}

This inequality is important, because it defines the operational
limit in taking continuous phase space measurements using coherent
states. The change $\delta z$ in the argument of the coherent
state  cannot be made that small (even if it is only in the $q$
direction) as to make
 meaningful the procedure of taking the continuum limit.
There exists a  minimum length on phase space given by $2 \Delta p
\Delta q + C_{pq}$  and any phase space process
 we monitor consists of discrete steps of this order of magnitude. This minimum
  length depends on the choice of coherent states, in other words on the choice
   of the filters we use in order to describe phase space properties.
  Irrespective of the choice made the precision cannot be rendered smaller
than one.

We note here that we can also minimise the phase space distance
with respect to all possible choices of coherent states. For this
purpose we use equation (3.13) to get
\begin{eqnarray}
\delta s^2 \geq (\Delta p)^2 \delta q^2 + (\Delta q)^2 \delta p^2 - 2 |C_{pq}|
\delta p \delta q \geq
\\ \nonumber
(\Delta p)^2 \delta q^2 + (\Delta q)^2 \delta p^2 - \sqrt{ (\Delta q)^2
(\Delta p)^2 - \frac{1}{4}}.
\end{eqnarray}
The right-hand side has a maximum, when the term in the square
root vanishes. In this  case,
\begin{eqnarray}
\delta s^2 \geq \frac{ \delta q^2}{4 (\Delta q)^2} + (\Delta q)^2 \delta p^2 \geq
\delta q \delta p
\end{eqnarray}

This result makes it clear that {\em the relation $\delta s^2 \geq
1$ is equivalent to Heisenberg's uncertainty relation}. Moreover
since $\delta s^2$ has a lower bound, the corresponding term in
the decoherence functional cannot be made to vanish, at least for
the case of measurements. Hence, the quantum anti-Zeno effect (or
the quantum Zeno effect) cannot really arise in a concrete
measurement situation;  see  also a similar analysis in
\cite{GOWR79}.

\subsubsection{Similarity to a diffusion process}

In taking the continuum limit, we have assumed that the paths
$z(\cdot)$ are continuous functions of a time parameter $t$. In
standard treatments this parameter is identified with Newtonian
time. In that case $\delta z \sim \delta t$ and $\delta s^2 \sim
\delta t^2$ and the term with the metric in the decoherence
functional (3.8) goes to zero.

However, in equation (3.8)  the specification of the time
parameter appears  nowhere as we assumed a vanishing Hamiltonian.
 The decoherence functional and probabilities are written solely in terms of the
$\delta z$. But this does not mean that the notion of time is
lost. Recall the distinction of the two properties of time: its
function as a  causal ordering parameter and its function as the
parameter   measuring evolution are distinct \cite{Sav99a}.
 Even with vanishing Hamiltonian the causal ordering is still present: it
  is the sequence by which the various single time properties
   $z_i$ enter in order to form a path. The ordering structure does not
  depend on the numerical value of the time parameter;
 it suffices to designate which property is first
 (or if one prefers which measurement takes place first.)
 There is no {\em a priori} reason to consider external Newtonian time
 as the parameter determining the path.

More precisely, the uncertainty relation $\delta s^2 \geq 1$
strongly suggests that the paths $z(\cdot)$ cannot be taken as
continuous (or at least differentiable) functions of the time $t$,
for in this case $\delta s^2$ could be made arbitrarily small by
taking $\delta t $ going to zero.

Let us consider now an (N+1)-time fine-grained history $\alpha =
(z_0, z_1, z_2, \ldots z_N)$. If we write $\delta z_i = z_{i} -
z_{i-1}$, the probability for $\alpha$ is
\begin{equation}
p(\alpha) = e^{- \sum_i \delta s_i^2} \geq e^{- \sum_i 1} = e^{-N},
\end{equation}
where the inequality is due to the uncertainty relation.
 This seems  like a nice behaviour: the more
 single-time measurements involved in the monitoring of a path,
 the smaller the probability that we are going to
 get it right.

 On the other hand, the behaviour of the limiting probability for  a path going
 with $e^{-N}$ is similar to that of classical decay process.
  We must note here that some caution is needed in the drawing of
  any quantitative conclusions from such
analogies, since the quantum probabilities (3.19) do
  not correspond to an additive probability measure. Nonetheless, the analogy helps
to demonstrate some interesting  points.

Let us consider a Wiener process on a Riemannian manifold. This is
associated to the heat equation for the single-time probability
distribution
\begin{equation}
\frac{\partial}{\partial t} \rho = D \nabla^2 \rho,
\end{equation}
with $D$ a diffusion constant. The probability for  a small
transition
\begin{equation}
p(\alpha) = e^{-\frac{D}{2 \delta t} \delta s^2}.
\end{equation}
Unlike (3.9) this expression is exact for all values of $\delta t$
and $\delta z$.

In a Wiener process the  time $t$ appears explicitly. Then one can
define the derivative $\dot{z}$   with respect to $t$ and write
the probability distribution for
 a path
\begin{equation}
p(z(\cdot)) \sim e^{-\frac{D}{2} \int dt g_{ij}(z) \dot{z}^i \dot{z}^j}.
\end{equation}
This expression is formal: one needs to  specify the discrete
version of the integral, because most paths in which the measure
is defined are not differentiable \footnote{This is in analogy
with stochastic processes and it is the reason the terms with the
metric, which are of order $(\delta z)^2$ cannot be ignored in the
continuum limit.}. However, these
 issues can be ignored since we are only interested in the  formal comparison
 of (3.24) with the quantum mechanical equation (3.9)

Clearly the main difference between the two equations is the presence of the term
 $\delta t$ in the denominator of the exponent of the probability. This signifies the
appearance of Newtonian time in the probability measure and
guarantees its finiteness.

To follow this analogy in the quantum case we  introduce a
time-step $1/\nu$, which corresponds to the duration of a single
time-step in terms of Newtonian time.  For timescales much larger
than $\nu^{-1}$ we can then approximate $z(t)$ by  a continuous
path. In this case $\delta z^i = \frac{1}{\nu} \dot{z}^i $ and
 the probability  reads
\begin{equation}
p(z(\cdot)) \sim e^{-\frac{1}{\nu} \int_0^t ds g_{ij}(z) \dot{z}^i
\dot{z}^j}.
\end{equation}
Comparing with (3.13) we  see that $\nu^{-1}$ is analogous to a
diffusion coefficient.

In the same approximation the decoherence functional (3.8) reads
\begin{equation}
d(\alpha, \beta) = \exp \left( i \int_C A - \frac{1}{2 \nu}
\int_C ds g_{ij}(z) \dot{z}^i \dot{z}^j \right).
\end{equation}
This expression converges to the Berry phase for $\nu \rightarrow
\infty$. In order to avoid any misunderstanding, we should remark
that the expression (3.24) for the decoherence functional is
completely different and leads to different properties from the
ones arising in studies of decoherence in open quantum systems.

The parameter $\nu$  was first introduced by Klauder as a means of
regularising the coherent state path integral in a geometrical
fashion \cite{Kla88}. He introduced $\nu$ as a diffusion
coefficient of a Wiener process on phase space, which is
eventually taken  to infinity. He proved that the coherent state
propagator equals
\begin{equation}
 \langle z|e^{-i\hat{H}t}|z' \rangle =
 \lim_{\nu \rightarrow \infty} \int Dz(\cdot) e^{\nu t} e^{i \int A
 - i \int_0^t ds H  -
 \frac{1}{2 \nu} \int_0^t ds g_{ij} \dot{z}^i \dot{z}^j }.
\end{equation}

However, if $\nu$ is not taken to infinity, the propagation is not
unitary. Klauder and
 Maraner explored this case and they showed that
 in the deterministic regime characterised by the
saddle-point approximation of (3.27)  many features of quantum
theory can
 be recovered \cite{KlMa97}. They commented, however, that if one  takes this picture
   seriously, one would have
   to deal with the constraints of Bell's theorem. However, this is not problematic
   when one
   does not seek a deterministic theory on phase space: the description
    by means of the decoherence functional employs non-additive probabilities
and is, therefore, not constrained by Bell's theorem.

\subsection{Introducing dynamics: time parameterisation}

\subsubsection{ The extended phase space}

In our work so far, the external Newtonian time did not appear
explicitly in the decoherence functional. This was due to the
absence of the Hamiltonian, which is the generator of translations
in Newtonian time.

 If we take the Hamiltonian operator into account the kernel
$\langle z|e^{-i\hat{H}(t-t')}|z' \rangle$  appears in the
expression (2.7) for the  decoherence functional. It is convenient
to define the new coherent states
\begin{equation}
|z, t\rangle = e^{-i \hat{H}t} |z \rangle.
\end{equation}
Here the vector $|z,t \rangle$ is parameterised by elements of
$\Gamma \times {\bf R}$. The vectors $|z \rangle$  will still form
coherent states in the generalised sense, as they provide a map
from $\Gamma \times {\bf R}$ to the projective Hilbert space. In
this case there does not exist a decomposition of the unity, but
one can still pullback the metric and connection to $\Gamma \times
{\bf R}$.

In what follows we shall write  $Z= (z,t)$ and  use Greek letters
for the indices of the tensors on $\Gamma \times{\bf R}$, so that
$Z^0 :=t$ and $Z^i := z^i$.  It is convenient to define the
operators $\hat{A}^{\mu}$. Explicitly
\begin{equation}
\hat{A}_0 = \hat{H},
\end{equation}
 and $\hat{A}_i$ is defined by its action on coherent states
\begin{equation}
\hat{A_i} |z \rangle = -i |\partial_i z \rangle .
\end{equation}
We denote tensors on $\Gamma \times {\bf R}$ with an overbar. In
this notation
\begin{eqnarray}
\bar{A}_{\mu}(Z) = \langle Z|\hat{A}_{\mu}|Z \rangle \\
\bar{g}_{\mu \nu} = \langle Z|\hat{A}_{\mu} \hat{A}_{\nu}|\bar{Z} \rangle -
\langle Z|\hat{A}_{\mu}|Z \rangle \langle Z| \hat{A}_{\nu}|Z \rangle
\end{eqnarray}

On $\Gamma \times {\bf R}$  the two form
\begin{equation}
\bar{\Omega} = d\bar{A} =\Omega - dH(z) \wedge dt
\end{equation}
 is degenerate. This means that $\Gamma \times {\bf R}$ is a presymplectic
manifold and  $\Gamma$ is obtained  by excising the degenerate
directions of $\Omega$.

Alternatively, this process can be described in terms of the
theory of constraints. To see this consider the manifold $\Gamma
\times {\bf R}^2$ with points $(z,t,p_t)$. The two-form
 $\Omega + dp_t \wedge dt$ is non-degenerate, thus
 $\Gamma \times {\bf R}^2$ is a symplectic manifold. If we
 consider the first-class constraint
\begin{equation}
\phi(z,t,p_t) = p_t + H(z) = 0,
\end{equation}
 then $\Gamma \times {\bf R}$ together
 with the two-form (3.30) is the constraint surface. We can obtain
 $\Gamma$ by the standard procedure of symplectic reduction, i.e. $\Gamma$ is defined
 as the set of all orbits on the constraint
 surface under the symplectic transformations generated by the constraint.
  First-class constrained  systems with this behaviour are called parameterised and
have many common features with general relativity.

Now, we can write the decoherence functional in complete analogy
to (.14)

\begin{equation}
d(\alpha, \beta) = \exp \left( i \sum_a \bar{A}(Z_a + \frac{\delta
Z_a}{2}) \delta z_a - \frac{1}{2} \sum_a \delta \bar{s}_a^2
\right) + O(\delta z^3).
\end{equation}

If we  consider continuous paths we obtain
\begin{equation}
d(\alpha, \beta) = e^{i \int_C\bar{A}} = e^{i \int_C (A - H dt)} = e^{i S},
\end{equation}
which means that the decoherence functional equals the value of
the phase space action along the
 closed loop formed by the two histories.

We should present here an important point of the history
formalism. While we could in principle study any path on
 $\Gamma \times {\bf R} $, physical histories are only the ones for which the time
parameter increases along the {\em physical temporal ordering}.
This means that if we have a history $\alpha = (Z_1, Z_2,
\ldots,Z_n)$ it cannot be physical unless  $t(Z_1) < t(Z_2) <
\ldots t(Z_n)$. If this is not the case then any predictions will
be nonsensical as the time ordering enters crucially in the
definition of the
 decoherence functional. The exact values
  of the time parameter do not matter as long as the ordering is preserved.
This is another  reflection of the distinction between the
ordering and the evolutionary properties of time, which
characterises any history
 description \cite{Sav99a}. For example,
 if we work on a parameterised system we need to specify a
time-ordering in the space of all possible histories
\cite{Har96,SavAn00}.  In the above case time is assumed Newtonian
and the ordering is trivial; it could be more interesting in a
consideration of relativistic systems.

\subsubsection{Time-energy uncertainty relation}

Once more we engage in the   discussion of the uncertainty
relation. The metric on the extended phase space reads
\begin{equation}
\delta \bar{s}^2 = \delta s^2(t) + 2 C_{EA_i} \delta z^i \delta t
+ (\Delta E)^2 \delta t^2.
\end{equation}
Here $C_{EA_i}$ is the correlation between
the Hamiltonian and the operator $\hat{A}_i(t) =
e^{i \hat{H}t} \hat{A}_i e^{- \hat{H}t}$, i.e.
\begin{equation}
C_{EA_i} = \frac{1}{2} \langle z|\hat{A}_i(t) \hat{H} + \hat{H}
\hat{A}_i(t) |z\rangle -  \langle Z|\hat{A}_i(t)|Z \rangle \langle
Z|\hat{H}|Z \rangle,
\end{equation}
while $(\Delta E)^2$ is the energy uncertainty on the coherent
state $|z \rangle$ \footnote{ Note that $\delta \bar{s}^2(t) =
(\Delta \delta \hat{S})$, where $\delta \hat{S} = \hat{A}_i(t)
\delta z^i - \hat{H} \delta t$ is the operator that measures the
change of action corresponding to the the transition of $\delta
z^i$ and $\delta t$. }.

In order to simplify the expressions, we define the operators

\begin{eqnarray}
\hat{C} = \Delta \hat{A}_i(t) \delta z^i = \hat{A}_i(t) \delta
z^i - \langle Z|\hat{A}_i(t)|Z \rangle \delta z^i
\\
\hat{D} = -   \Delta \hat{H} \delta t = - (\hat{H} \delta t -
\langle Z|\hat{H}|Z \rangle \delta t).
\end{eqnarray}
We then write the following inequality
\begin{equation}
\langle \psi| \hat{C}^2| \psi \rangle \langle \psi|\hat{D}^2| \psi \rangle -
\frac{1}{4} (\langle \psi|\hat{C} \hat{D} + \hat{D} \hat{C}| \psi \rangle )^2 \geq
\frac{1}{4} | \langle \psi|[\hat{C}, \hat{D}]| \psi  \rangle|^2 ,
\end{equation}
for which inequality (3.15) is a special case \footnote{Explicit
writing down of all terms in (3.41) will convince the reader
 that this  is a consequence of Schwartz's inequality.}.
Now
\begin{eqnarray}
[\hat{C}, \hat{D}] = - [\hat{A}_i(t), \hat{H}] \delta z^i \delta t
=  - e^{i \hat{H}t } [\hat{A}_i \delta z^i, \hat{H}] e^{-i \hat{H}t} \delta t.
\end{eqnarray}
The commutator equals  $\frac{\partial}{\partial z^i} \hat{H}
\delta z^i$, which  in turn equals  $\delta \hat{H}$, i.e.  the
change in the Hamiltonian by virtue of a translation by $\delta
z^i$ on phase space.

If we define
\begin{equation}
\delta H(z,t) = \langle z| e^{i \hat{H}t} \delta \hat{H} e^{-i \hat{H}t}| z \rangle,
\end{equation}
we obtain
\begin{equation}
[ \hat{C}, \hat{D}] = - \delta H(t,z) \delta t.
\end{equation}
The inequality (3.41) implies that (after we drop  $\psi $ for
simplicity)
\begin{equation}
|\langle \hat{C} \hat{D} + \hat{D} \hat{C} \rangle|
 \leq \sqrt{4 \langle C^2 \rangle \langle D^2 \rangle - \delta H^2 \delta
 t^2},
\end{equation}
hence,
\begin{equation}
\delta \bar{s}^2 \geq \langle \hat{C}^2 \rangle + \langle \hat{D}^2 \rangle -
 \sqrt{4 \langle C^2 \rangle \langle D^2 \rangle - \delta H^2 \delta t^2}.
\end{equation}
Now the right-hand-side takes maximum value when the term in the
square root
 vanishes,
\begin{equation}
4 \langle C^2 \rangle \langle D^2 \rangle = \delta H^2 \delta t^2.
\end{equation}

Then
\begin{equation}
\delta \bar{s}^2 \geq \langle \hat{C}^2 \rangle +
 \frac{\delta H^2 \delta t^2}{4 \langle C^2 \rangle }.
\end{equation}
We can minimise the right-hand-side of (3.48) with respect to
$\langle C^2 \rangle$ to get
\begin{equation}
\delta \bar{s}^2 \geq \delta H \delta t,
\end{equation}
where $\delta H$ is the difference in the classical value of the
energy between two specifications of phase space points.

Due to the uncertainty in the specification of phase space  points
there exists an uncertainty $\delta E$ in the specification of the
energy: $\delta H$ cannot be smaller  than this energy.  Taking
into account the time-energy uncertainty principle for $\delta E$
\footnote{The time-energy uncertainty relation exists only in one
form, the Heisenberg one. There does not exist a statistical form,
because quantum theory does not accept a physical time operator.
See \cite{Bus01}  for a recent review.}, namely   $\delta E \delta
t \sim 1$, we get
\begin{equation}
\delta \bar{s}^2 \geq 1.
\end{equation}

So far we have seen that the quantum evolution of a quantum system
with non-zero Hamiltonian is best described in terms of the
geometry of the extended phase space, which includes time as a
parameter. The condition for the Riemannian metric on
$\Gamma_{ext}$,  $\delta \bar{s}^2 \sim 1$, is equivalent to
Heisenberg 's uncertainty relation. Note that the  derivation of
this equation  did not specify the physical system's phase space.
Hence we conclude that the geometric description can be used to
implement Heisenberg's principle in {\em any physical system},
including systems that are described by topologically non-trivial
phase spaces.

Like in the case of vanishing Hamiltonian,  for a history $\alpha$
with N time-steps the probability is
\begin{equation}
p(\alpha) \sim e^{-N}.
\end{equation}

We can write a sharper inequality for this probability. If we
assume that $\delta H(t)> \delta E_t$, we can go to the continuous
limit, namely
\begin{equation}
p(\alpha) \sim e^{- \int dt \delta E_t},
 \end{equation}
It is important to remark that $\delta E_t$ is not proportional to
$\delta t$ as it corresponds to an irreducible spread of energy of
the quantum state at a moment of time. Equation (3.52) is very
interesting. It strongly suggests that the assignment of
probabilities to histories primarily depends on the time-averaged
energy uncertainty $\langle \delta E \rangle = \frac{1}{t}
\int_0^t ds \delta E_s ds$ for the paths:
 the most probable paths being characterised by the smallest values
 of $\langle \delta E \rangle$. Essentially $(\langle \delta E \rangle)^{-1}$
is the decay time of the probabilities for the continuous paths.
 The metric also contributes to
 the off-diagonal elements of the
decoherence functional. Paths characterised by large values of
time-averaged energy uncertainty seem to decohere more
efficiently.

\section{An example: spinless relativistic particle}

\subsection{The coherent states}
As a non-trivial application of our previous ideas we  study the
phase space geometry of a relativistic particle with zero spin.
The relevant symmetry group  is the Poincar\'e group, so we
construct its associated coherent states.

We recall here that the Poincar\'e group is the semidirect product
of the Lorentz group with the Abelian group of spacetime
translations. Its representations are characterised by the value
of the mass $m$ and spin $s$. In this paper we shall study the
case  $m \neq 0, s =0  $, namely a massive spinless  particle.

We consider the space of all unit, timelike vectors $\xi$, with
positive values of the zero-th component, $V = \{\xi| \xi_{\mu}
\xi^{\mu} = 1, \xi^0 \geq 0 \}$. The space $V$ carries a
Poincar\'e invariant measure
\begin{equation}
d \mu(\xi) = m^2 \frac{d^3 {\bf \xi}}{2 \omega_{\xi}},
\end{equation}
where $\omega_{\xi} = \xi^0 = \sqrt{1 + {\bf \xi}^2}$.

The Hilbert space on which the representation is constructed is
${\cal L}^2(V, d \mu)$ and the Poincar\'e group action \footnote{
Let us denote by $\hat{P}^{\mu}$ and $\hat{M}^{\mu \nu}$  the
generators of the translations and the Lorentz group respectively.
We can distinguish the boost generators $\hat{K}^i = \hat{M}^{0i}$
from the rotation ones $\hat{N}^i = \frac{1}{2} \epsilon^{ijk}
\hat{M}_{ij}$, by making reference to a timelike direction. The
generators read explicitly $\hat{P}^{\mu} = \xi^{\mu}$, $\hat{{\bf
N}} = - i {\bf \xi} \times {\bf \nabla_{\xi}}$ and $ \hat{{\bf K}}
=  -i \xi^0 {\bf \nabla_{\xi}}$, where $(\nabla_{\xi})_i =
\frac{\partial}{\partial \xi^i} + \frac{\xi^i}{\xi^0}
\frac{\partial}{\partial \xi^0}.$ }
\begin{eqnarray}
\hat{U}(\Lambda) \Psi(\xi) =  \Psi(\Lambda^{-1} \xi), \\
\hat{U}(X) \Psi(\xi) = e^{i m \xi \cdot X} \Psi(\xi).
\end{eqnarray}

In order to construct a set of coherent states we need to choose a
reference vector $\Psi_0$, which is invariant under the action of
the maximal compact subgroup of the Poincar\'e group, namely the
group $SO(3)$ of spatial rotations. This suggests that the
reference vector  depends on $\xi$ only through the product
$n_{\mu} \xi^{\mu}$, where $n_{\mu}$ is a unit timelike vector.

We choose  a Gaussian reference vector
\begin{equation}
\Psi_0(\xi) = \frac{1}{m (\pi \sigma^2)^{3/2}} (2 n \cdot
\xi)^{1/2}  e^{- \frac{1}{2 \sigma^2} \xi \cdot {}^n\Gamma \cdot
\xi},
\end{equation}
where ${}^n\Gamma_{\mu \nu} = -\eta_{\mu \nu} + n_{\mu} n_{\nu}$.
This vector is centered around $\xi^i = 0 $ with a width equal to
$\sigma$.

For fixed $n$ we can distinguish the elements of the Lorentz group
into the ones that leave $n$ invariant (which form a subgroup
SO(3) of spatial rotations) and the ones that do not. The latter
generate boosts and can be parameterised by a unit timelike vector
$I$, such that
\begin{equation}
\Lambda_I n = I.
\end{equation}
The above equation defines an isomorphism, because for each $I$
there exists a unique $\Lambda_I$ satisfying (4.9) and vice versa.
Any Lorentz matrix can be written as a product $\Lambda = R
\Lambda_I$ for some rotation matrix and vector $I$. When we act on
our reference vector with $\Lambda$, the rotation matrix does not
contribute as it  leaves the product $n \cdot \xi $ invariant.
Hence only the boost part of the Lorentz matrix acts
non-trivially. Thus the coherent states associated to the
Poincar\'e group  depend on $I$ and the parameters $X$,  which
correspond to spacetime translations,
\begin{eqnarray}
\Psi_{X,I} (\xi) = \hat{U}(X) \hat{U}(\Lambda_I) \Psi_0(X) =
e^{i m \xi \cdot X} \Psi_0(\Lambda_I^{-1} \xi) =
\nonumber \\
\frac{1}{m (\pi \sigma^2)^{3/2}} (2 I \cdot \xi)^{1/2}  \exp
\left( -\frac{1}{2 \sigma^2} (I_{\mu} I_{\nu} - \eta_{\mu \nu})
\xi^{\mu} \xi^{\nu} + i m \xi_{\mu} X^{\mu} \right).
\end{eqnarray}

An interesting point is to examine the expectation value of the
energy-momentum vector on the coherent states
\begin{equation}
\langle X,I| \hat{P}^{\mu} |X,I \rangle = m \kappa I^{\mu},
\end{equation}
 where
\begin{equation}
\kappa = \int d \mu(\xi) \omega_{\xi} |\Psi_0|^2(\xi)
\end{equation}
is  the expectation value of the energy $n_{\mu} \hat{P}^{\mu}$ on
$\Psi_0$ divided by $m$.  Because  the coherent state is not
localised at a point of $V$ there exists  a spread in the
distribution of energy, even though the spatial momenta vanish. If
equation (4.7) is to conform with our expectations that the
expectation value of $\hat{P}^{\mu}$ is the momentum associated to
the coherent states, we need to redefine the mass on  phase space
as $M = \kappa m$. We shall later show that this is  the correct
account.

We can estimate $\kappa$

\begin{equation}
 \kappa = 1 + \frac{1}{4}  \sigma^2 - \frac{1}{16} \sigma^4 +
 O(\sigma^6).
\end{equation}

The set of coherent states (4.6) depends on seven parameters,
three corresponding to the momenta $I$ and four corresponding to
the spacetime translations. Clearly one of these parameters  plays
the role of time, hence these coherent states are similar to the
states $| z,t \rangle$ of the previous section. They are not
expected to have a decomposition of the unity.

Furthermore these states do not depend on the choice of $n$, even
though the reference vector did. The set of coherent states is
invariant under the action of the Poincar\'e group
\begin{eqnarray}
\hat{U}(\Lambda) |X,I \rangle = |\Lambda X, \Lambda I \rangle \\
\hat{U}(Y) |X,I \rangle = |X+Y,I \rangle .
\end{eqnarray}

We can calculate the fine-grained decoherence functional (3.35) in
terms of these coherent states.  Note again that in the
specification of any path the causality condition has to be
specified. Hence a phase space point $(X',I')$ is in the future of
another phase space point $(X,I)$ only if the causal structure of
Minkowski spacetime is respected. That is, $X'$ has to lie in the
causal future of $X$. This means that $X$ and $X'$ have to be
timelike or null separated and $X'^0 \geq X^0$. Furthermore, in
order  to describe physical particles we  demand that $I^0
> 0$.

We can  reduce the set of coherent states by taking a fixed value
of the parameter $t = n \cdot X$, i.e. treating $t$ as an external
parameter and not as an argument of the coherent states.

In this way we may define  our coherent states at an instant of
time,  a spacelike three-surface
  $\Sigma$, which is uniquely determined by the choices of $n$ and $t$.
  The coherent states then
  depend on the spatial variables $x$ and $I$, which are the projections of
  $X$ and $I$ on $\Sigma$ and they span
    the phase space of a single particle $T^*\Sigma$.

We  denote the coherent states restricted on $\Sigma$ as $|x, I
\rangle_{\Sigma}$. The Poincar\'e group behaves as follows:
transformations that leave $\Sigma$ invariant (spatial rotations
and translations) preserve the coherent states, while the ones
that take $\Sigma$ to  another surface $\Sigma'$ (namely boosts
and time translations) also take the set of coherent states into
the  one associated to $\Sigma'$.

For the restricted coherent states we can calculate
\begin{equation}
\int d^3 I d^3 x \langle \xi | x, I \rangle_{\Sigma} {}_{\Sigma}
\rangle x, I| \xi' | = \frac{1}{m^3} 2 \kappa  \omega_{\xi}
\delta^3(\xi - \xi'),
\end{equation}
and thus implies the existence of a resolution of the unity
\begin{equation}
 \kappa \hat{1} =  m^3 \int d^3I d^3x  |x, I \rangle_{\Sigma} {}_{\Sigma} \langle x, I| .
\end{equation}

In what follows, we shall denote the measure $m^3 d^3I d^3x$ as $d
\mu_{\Sigma}(I,x)$.

Since  the decomposition of the unity is necessary in order to
define the operators that represent physical quantities, any such
definition necessarily depends on the choice of the surface
$\Sigma$. For example, this is the case of the position operator
\begin{equation}
\hat{{\bf x}}_{\Sigma} = \frac{1}{\kappa}
 \int d \mu_{\Sigma}(I,x) {\bf x} |x,I \rangle_{\Sigma} {}_{\Sigma} \langle x,I|,
\end{equation}
which coincides with the one defined by Newton and Wigner
\cite{NeWi49}.

Hence, even if the fine-grained decoherence functional is a fully
covariant object, the specification  of any correlation function
makes reference to some spacelike surface $\Sigma$ and hence
breaks the covariance. This is also true for the  coarse-grained
histories. Coarse-graining over phase space involves integration
with the measure $d \mu_{\Sigma}$ and hence refers to a spacelike
surface. Hence any prediction of the theory carries implicitly a
reference to a chosen hypersurface. This is already known in
canonical quantisation and is highlighted in the histories
formalism \cite{Sav01a}.

\subsection{Phase space  geometry}

The first step towards determining the geometric objects  on
$\Gamma$ is to compute $d| X,I \rangle$. It equals
\begin{equation}
d \Psi_{IX}(\xi) = \left [\frac{\xi \cdot d I}{2 I \cdot \xi} -
\frac{\xi \cdot I \xi \cdot dI}{\sigma^2} + i m \xi \cdot dX
\right] \Psi_{IX}(\xi),
\end{equation}

from which we obtain
\begin{equation}
\langle X,I|d|X,I \rangle = i m \kappa I^{\mu} d X_{\mu} +
\frac{1}{2}(1 - \frac{\kappa}{\sigma^2}) I^{\mu} dI_{\mu}.
\end{equation}
The fact that $I^2 = 1$ implies that $I_{\mu} dI^{\mu} = 0$, hence
\begin{eqnarray}
A = \kappa m I^{\mu} dX_{\mu}, \\
\Omega = \kappa m dI^{\mu} \wedge dX_{\mu}.
\end{eqnarray}

It is clear from the above that the correct definition of the
classical momentum is $P^{\mu} = m \kappa I^{\mu}$. However, this
also implies that $P_{\mu}P^{\mu} = \kappa^2 m^2$ \footnote{ In
quantum theory the Wigner representation theory says that the
spinless representations of the Poincar\'e group are classified by
the value of the parameter $m = \hat{P}^{\mu} \hat{P}_{\mu}$.In
symplectic mechanics the theory of Konstant-Souriau states that
the symplectic actions of the Poincar\'e group (in the spinless
case) is classified by the value of $P_{\mu} P^{\mu} = M^2$. The
coherent states provide a map between quantum theories
corresponding to parameter $m$ and classical symplectic manifolds
characterised by the value $M$. There is no {\em a priori} reason
why these parameters should have the same value.}.

The calculation of the metric is straightforward but tedious. The end result is
\begin{eqnarray}
ds^2 = - \frac{\alpha}{3 \sigma^2} \eta_{\mu \nu}  dI^{\mu}
dI^{\nu} +  K_{\mu \nu} dX^{\mu} dX^{\nu}.
\end{eqnarray}
The first term is the Riemannian metric on $V$  inherited from the
Lorentzian metric on Minkowski spacetime times a constant. The
parameter $\alpha$ equals
\begin{equation}
\alpha = \frac{1}{(\pi \sigma^2)^{1/2}}  \int_0^{\infty} \frac{d
\xi}{1 + \xi^2} e^{- \xi^2/ \sigma^2} = 1 + O(\sigma^2).
\end{equation}

The second term is characterised by the tensor
\begin{equation}
K^{\mu \nu} = \langle X,I|\hat{P}^{\mu} P^{\nu} |X,I \rangle -
\langle X,I|\hat{P}^{\mu} | X,I\rangle \langle X,I|\hat{P}^{\nu} |
X,I\rangle,
\end{equation}
which is the correlation tensor for the four-momentum on a
coherent state. Explicitly,
\begin{equation}
K_{\mu \nu} = m^2[(1 + \frac{2}{3}  \sigma^2 -
\kappa^2)I_{\mu}I_{\nu} -\frac{1}{6}  \sigma^2 \eta_{\mu \nu}].
\end{equation}

We want the coherent state to be very sharply peaked around
$\xi^{\mu} = I^{\mu}$, so that it will correspond as closely as
possible to a phase space point. For this purpose we take $\sigma
<< 1$. In this case the dominant terms are
\begin{equation}
\delta \tilde{s}^2 = \left( \frac{1}{3 \sigma^2}  + O(\sigma^0)
\right) \eta_{\mu \nu} \delta I^{\mu} \delta I^{\nu} + m^2 \left(
\frac{\sigma^2}{6} + O( \sigma^4) \right) (I_{\mu} I_{\nu} -
\eta_{\mu \nu} ) \delta X^{\mu} \delta X^{\nu}.
\end{equation}

If we minimise the right-hand-side with respect to $\sigma^2$ we
get
\begin{equation}
\delta \tilde{s}^2 \geq m \frac{\sqrt{2}}{3 } \delta I \delta_I X,
\end{equation}
where
\begin{eqnarray}
\delta I = \sqrt{ - \eta_{\mu \nu} \delta I^{\mu} \delta I^{\nu} } =
\sqrt{(\delta_{ij} - \frac{I_i I_j}{1 + I^2}) \delta I^i \delta I^j}\\
\delta_IX = \sqrt{ (I_{\mu} I_{\nu} - \eta_{\mu \nu} ) \delta X^{\mu} \delta X^{\nu} }.
\end{eqnarray}
The uncertainty relation $\delta \tilde{s}^2 \sim 1$ implies  that
\begin{equation}
 m \delta I \delta_IX \geq 1 .
\end{equation}
Equation (4.24) provides a covariant generalisation of the
time-energy uncertainty relation.

 In the particle's rest frame we have  $I^{\mu} = n^{\mu}$. Therefore
 the condition $I^{\mu} \delta I_{\mu} = 0$ implies that $\delta I^0 = 0$, so
that
\begin{eqnarray}
\delta I = \sqrt{\delta {\bf I} \cdot \delta {\bf I}}, \\
\delta_IX = \sqrt{ \delta {\bf x} \cdot \delta {\bf x}},
\end{eqnarray}
which means that (4.31)  coincides with the non-relativistic
uncertainty relation.

Of special interest is  the degenerate case that $\delta X^{\mu} =
\delta t I^{\mu}$. This corresponds to the case that the measuring
devices are clocks along the classical trajectory of the particle.
In this case the dominant term of $K_{\mu \nu}$ vanishes and we
have to use the next term in the expansion of $\kappa$. This
yields
\begin{equation}
\delta \bar{s}^2 = \frac{m^2}{3 \sigma^2} (\delta I)^2  +
\frac{m^2 \sigma^4}{16} (\delta t)^2.
\end{equation}
Minimising with respect to $\sigma$ we get
\begin{equation}
\delta \bar{s}^2 \geq \frac{3^{1/3}}{4} m^{2/3}  (\delta I)^{4/3} (\delta t)^{2/3}.
\end{equation}

This implies that the  Heisenberg uncertainty relations take a
rather unusual form
\begin{equation}
m^{2/3} (\delta t)^{2/3} (\delta I)^{4/3} \sim 1.
\end{equation}

\section{Conclusions}
We have studied the role that the phase space geometry plays in
the probability assignment for continuous-time histories. In
particular, we showed that the coherent states induce a metric on
the phase space, which proves to be a key ingredient of the
decoherence functional.

This metric has a significant physical interpretation as it
provides a geometric way of formulating the Heisenberg uncertainty
relations. This version of the uncertainty relations can be
implemented in any classical phase space, including the ones with
non-trivial topological structure. An interesting byproduct of our
results is a new proof of the impossibility of the physical
manifestation of the quantum Zeno and anti-Zeno paradoxes.

It is important to remark that in the generic case, that the
Hamiltonian does not vanish, the physically relevant metric is
defined on the extended phase space, which includes time as a
parameter. Since the decoherence functional is constructed from
the metric and the phase space action, our construction provides a
stepping stone for a geometric quantisation algorithm for
histories, which can be naturally generalised to include
parameterised systems, such as the relativistic particle we
considered in section 5. For this reason it might prove relevant
in the histories quantisation of general relativity \cite{Sav03}.
In fact, the search for a geometric procedure for the quantisation
of histories has been one of the main motivations of this work.

Finally, we note that our construction of relativistic coherent
states and the ensuing uncertainty relations are novel objects,
which demonstrate the versatility of our approach for the study of
the quantum properties of a large class of physical systems.

\bigskip

\noindent{\large\bf Acknowledgements}

\noindent We gratefully acknowledge support from the EPSRC
GR/R36572 grant (N.S.) and from the Marie Curie Fellowship of the
European Commission (C.A).

\end{document}